# Integration of Programmable Diffraction with Digital Neural Networks

Md Sadman Sakib Rahman[1,2,3]     mssr@ucla.edu
Aydogan Ozcan[1,2,3,*]             ozcan@ucla.edu
1 Electrical and Computer Engineering Department, University of California, Los Angeles, CA, 90095, USA
2 Bioengineering Department, University of California, Los Angeles, CA, 90095, USA
3 California NanoSystems Institute (CNSI), University of California, Los Angeles, CA, 90095, USA
*Corresponding author: ozcan@ucla.edu

## Abstract
Optical imaging and sensing systems based on diffractive elements have seen massive advances over the last several decades. Earlier generations of diffractive optical processors were, in general, designed to deliver information to an independent system that was separately optimized, primarily driven by human vision or perception. With the recent advances in deep learning and digital neural networks, there have been efforts to establish diffractive processors that are jointly optimized with digital neural networks serving as their back-end. These jointly optimized hybrid (optical+digital) processors establish a new "diffractive language" between input electromagnetic waves that carry analog information and neural networks that process the digitized information at the back-end, providing the best of both worlds. Such hybrid designs can process spatially and temporally coherent, partially coherent, or incoherent input waves, providing universal coverage for any spatially varying set of point spread functions that can be optimized for a given task, executed in collaboration with digital neural networks. In this article, we highlight the utility of this exciting collaboration between *engineered and programmed diffraction* and *digital neural networks* for a diverse range of applications. We survey some of the major innovations enabled by the push-pull relationship between analog wave processing and digital neural networks, also covering the significant benefits that could be reaped through the synergy between these two complementary paradigms.

## Introduction
Deep learning has transformed various branches of scientific endeavors, such as computer vision, natural language processing, medical image analysis, bioinformatics, drug discovery, and material science, among many others[1]. This success of deep learning has been expedited by the exponential increase of data resulting from the proliferation of data-generating devices and sensors, such as smartphones, mobile devices, satellites, and social media platforms[2]. Powered by deep learning, the rise of generative artificial intelligence (AI) has encroached on areas previously thought to be the prerogative of humans only, such as the synthesis of music, artwork, and literature.[3] In the same vein, deep learning has also transformed the design of physical systems by breaking the limitations of human intuition and intruding into nonintuitive and better-performing design spaces.

Optical imaging and sensing are among the disciplines that have benefitted greatly from the progress in deep learning. Imaging and sensing systems, in general, rely on the engineering of light diffraction to carry information from the object of interest to the sensor plane. In most of the engineered diffractive systems, the information delivered to the detector array could be interpreted directly by humans. However, for autonomous systems enabled by machine vision, the need for human interpretability is lifted as long as the captured signals can be processed digitally by a computer to recover the desired information. Even if human interpretability is ultimately intended, an intermediate digital processing step complementing the analog diffractive processor often improves the performance of the overall system. As a result, instead of human perception-oriented optimization of diffractive elements, the focus shifted to the optimization of diffractive elements with the subsequent digital processing in consideration.

Broadly speaking, the augmentation of front-end optics with back-end digital processing for recovering information that would otherwise be lost or compromised sets one of the core pillars of computational imaging and sensing. Prior to the popularity of deep learning, the back-end digital processing algorithms were, in general, based on a model of the front-end physical processes. Data-driven deep learning models such as artificial neural networks (ANNs) have shown remarkable success in outperforming these model-driven algorithms, both in terms of performance and speed[4]. The joint optimization of the optical and digital processers, powered by deep learning, enables the exploration of design spaces that blend the best of both paradigms. Such optimization often leads to optics that integrate an enriched set of spatially varying point spread functions (PSFs), capable of fully exploiting the design degrees of freedom. Therefore, hybrid processing of optical information through the collaboration between engineered diffraction and digital ANNs has been undertaken for a variety of imaging and sensing tasks.

In this article, we shed light on the utility of this collaboration between engineered diffraction and digital ANNs in enabling new capabilities in computational imaging and sensing systems, dynamic metasurfaces, ANN acceleration, and optical-electronic encoding and decoding of information. Following introductory discussions on ANNs and free-space optical information processors in the next two sections, we will survey these diverse applications enabled by the collaboration between engineered diffraction and digital ANNs. The following points delineate the scope of this article:

(i) We focus our discussions on recent endeavors that utilize engineered diffraction with free-space optics. Therefore, we exclusively cover scenarios where the input analog information of interest can be directly accessed by free-space optics, avoiding additional steps such as image preprocessing, phase retrieval or vectorization of 2D or 3D information that would be necessary for, e.g., waveguide-based optical processors[5–11], which are excluded from our coverage.

(ii) This article focuses on the works where ANNs function collaboratively with a jointly optimized optical system to perform the task at hand. We have accordingly excluded a large body of work where ANNs have solely been utilized for the design of the optical system[12–17]; in these approaches, once the optimization/design is complete, the ANNs no longer play a role in performing the desired task, which is outside the scope of this article.

(iii) Related to computational imaging and related applications, we only cover works utilizing ANNs for back-end digital processing, where both optical diffraction and the corresponding digital ANN were jointly designed and optimized. Excluded from consideration in our article are, e.g., computational imaging and compressive sensing systems utilizing ANN back-ends with random masks or diffusers at the front-end, without any joint optimization process.

## Deep Learning and Artificial Neural Networks

Deep learning is a subfield of machine learning that focuses on training ANNs to capture patterns from data. Inspired by the structure and function of the human brain, deep ANNs are composed of multiple interconnected layers of computational nodes known as artificial *neurons*. In our article, we reserve the term 'ANN' for neural networks implemented on digital electronic platforms, which are to be distinguished from some alternative implementations reported in recent years. ANNs extract essential features from the input data by passing them through consecutive layers. The data-driven training of the deep layers in ANNs is made possible through algorithms such as error backpropagation and gradient descent[1]. In fully connected networks, also known as dense networks, each neuron in a layer is connected to every neuron in the subsequent layer. This interconnectedness enables fully connected networks to learn complex nonlinear mappings between their input and output layers, making them suitable for tasks such as classification and regression.

Convolutional neural networks (CNNs) form another key neural network architecture, particularly suited for processing spatially correlated data such as images. CNNs employ convolution operations to systematically extract hierarchical features from the image pixels. These convolutions are defined by learned filters that slide across the input feature map, giving CNNs their attributes, such as parameter sharing and local connectivity between neurons of consecutive layers. CNNs have revolutionized computer vision, where they are used for various tasks including, e.g., image recognition, object detection, and image segmentation, achieving state-of-the-art performance on various benchmarks[18].

'Universal approximation theorem' states that ANNs with a single hidden layer, given enough neurons, can approximate any continuous function to arbitrary accuracy[19]. This theorem underscores the remarkable expressive power of neural networks, highlighting their ability to represent highly complex functions. Nonlinear activation functions[19–23] that follow fully connected or convolutional layers of an ANN play a crucial role in bestowing neural networks their universal approximation capacity.

ANNs have also been successfully employed for various problems in optics and photonics, such as photonic inverse design[24,25], computational imaging and microscopy[4,26–29], sensing[30–35], and spectroscopy[36–45], among other applications[46,47]. However, a detailed discussion of these works is beyond the scope of this article since we do not cover the works where digital ANNs augment traditional optical systems or are only used in the design of the optics, as also outlined in our Introduction.

## Analog Information Processing with Diffractive Optics

Optics benefits from the large bandwidth and massive parallelism rising from different attributes of light, such as position, time, wavelength, and polarization, among others. At the same time, a computational task of interest can be completed via the propagation of optical fields through thin optical designs. For example, the phenomenon of free-space diffraction, i.e., merely the transformation of the optical field as it propagates, gives rise to a lowpass filtering operation with a pre-defined kernel. As another example, computing the Fourier transform of an optical field can be accomplished via light propagation through a lens[48]. Engineering diffraction in this manner also enables other kinds of linear operations. For example,

Fourier filtering by a desired filter is possible by placing an appropriately designed diffractive optical element (DOE) at the Fourier plane of a 4f system[48]. Such diffractive elements spatially modulate the amplitude or the phase of the incident light, and can be implemented by using transparencies or phase plates[49]. Reconfigurability of such spatial modulation of light is also possible by using a spatial light modulator (SLM)[50,51].

Another framework for engineering diffraction is formed by metasurfaces[52–55]. They belong to a broader category of materials known as metamaterials[56], which are artificially engineered with properties not found in naturally occurring materials. Engineered light-matter interactions through subwavelength structures enable metamaterials to exhibit extraordinary electromagnetic properties such as negative refraction. Metasurfaces, i.e., 2D metamaterials, comprise one layer or a few layers of repeated elements with subwavelength structures engineered to manipulate the properties of light. In analogy with atoms, these repeating building blocks of metasurfaces are called meta-atoms. By carefully designing the geometry, size, and arrangement of the subwavelength scatterers within these meta-atoms, metasurfaces can exert precise control over light propagation, enabling functionalities like flat lenses[57], beam steering[58], cloaking[59], and holography[60]. While both DOEs and metasurfaces can be used for wavefront modulation, they operate on different principles and offer distinct advantages[55]. DOEs rely on diffractive effects from varying thickness or refractive index, while metasurfaces utilize engineered nanostructures to control light at subwavelength scales, offering versatile optical manipulation in a thin, planar format.[61]

Diffractive deep neural networks ($D^2NNs$)[62], also known as diffractive networks or diffractive optical networks, constitute a recently emerging framework for diffraction engineering and deep learning-enabled optical computing. A diffractive optical network comprises a series of diffractive surfaces that are spatially engineered at a lateral scale of half a wavelength or larger, connecting the input and the output fields-of-view (FOVs) of an optical system. Data-driven supervised learning tools such as error backpropagation and stochastic gradient descent are used to digitally optimize the individual spatial features within these diffractive surfaces/layers. Once the digital training is complete, which is a one-time process, these surfaces can be fabricated and assembled to form a diffractive network, performing the desired computation all-optically at the speed of light propagation within a thin optical volume that would typically span < 100λ in its axial dimension. Some of these diffractive layers can also be implemented by SLMs or metasurfaces to introduce dynamic reconfigurability[63,64].

Just as deep learning provides the optimization tools for training multiple hidden layers of an ANN in a data-driven manner, it also enables the digital optimization of a set of diffractive surfaces forming a task-specific diffractive optical network, which is achieved through supervised learning with example pairs of input and target (ground truth) output optical patterns. The cascading of multiple surfaces within a deeper diffractive network architecture provides enhanced processing capability, better inference accuracy and improved output diffraction efficiency in comparison to a single surface that has the same number of degrees of freedom[62,65–70]. Despite relying on linear optical materials, $D^2NNs$ can be used to approximate nonlinear transformations[71–74], and have already been used to optically implement various tasks such as all-optical statistical inference, image encryption, all-optical phase recovery and phase conjugation as well as image reconstruction without a digital computer[62,71–73,75,76]. Given sufficient degrees of freedom in the form of optimizable features, diffractive networks can provide universal linear processing of spatially coherent and incoherent light, also covering broadband operation and wavelength multiplexing.[67,77–81] Furthermore, despite using isotropic materials as part of its diffractive surfaces, a D²NN design can also be programmed for universal polarization transformations, accurately approximating thousands of different spatially-encoded polarization scattering matrices at the diffraction limit of light[82].

## Integration of Programmable Diffraction and Digital ANNs

In this section, we cover the collaboration between engineered/programmable diffraction and digital ANNs for various applications, starting with computational imaging.

### Computational Imaging

Computational imaging[83] goes beyond a simple capture of the image of a specimen and involves digital processing of the raw data to generate images with e.g., enhanced performance or additional information that may not be directly observable with traditional imaging techniques. In general, computational imaging facilitates a wide range of capabilities such as super-resolution, volumetric imaging, hyperspectral imaging, extended depth-of-field and phase imaging, among others[84]. One of the techniques frequently used in computational imaging is the use of a coded aperture, i.e., a purposefully designed aperture pattern, which modulates the incoming light to encode the scene information into a limited number of measurements. By analyzing or decoding these raw measurements using computational algorithms, the original scene information can be digitally reconstructed. However, most of these image reconstruction algorithms are iterative and relatively slow, limiting the throughput of the imaging system. As an alternative, properly trained ANNs have been used for faster reconstruction of the object from the coded measurements in a single forward pass[85].

One strategy for designing an ANN-incorporated coded aperture-based computational imaging system is to follow the 'sequential' approach. In this case, a coded aperture is first handcrafted for effective encoding of the desired information, possibly guided by information theory-driven design principles. Subsequently, a reconstruction ANN is trained using the measurements captured through the designed coded aperture and their corresponding ground truths. Alternatively, since the measurement process through a coded aperture can, in general, be numerically modeled, the aperture encoding can be digitally formulated as the front-end of the reconstruction ANN. This digital embedding of the coded aperture within a deep ANN enables the 'joint' optimization of the coded aperture and the ANN[86], bringing together the best of both worlds, optical and digital. Such a data-driven optimization of a 'deeply coded aperture' is similar to the optimization of a diffractive optical network and leads to nonintuitive aperture designs that often deliver superior performance compared to intricate aperture designs based on information-theoretic principles. The term 'deep optics' has been coined to refer to such data-driven deep learning-based design of optical components[87].

Fueled by their initial success, deeply coded apertures have been used for various computational imaging tasks. One of the earlier works incorporating the joint optimization of the aperture code and the reconstruction algorithm is demonstrated by Shedligeri et al.[88] for the task of single-image depth estimation, see Fig. 1a. They formulated the reconstruction CNN to output the blur kernel size that encodes the depth information for the image patches. As for the coded aperture, they only considered binary amplitude modulation. Haim et al.[89] incorporated a phase-coded aperture with a jointly optimized CNN for depth estimation, resulting in higher light efficiency. Subsequent efforts included more sophisticated phase mask templates[90], the introduction of an occlusion-aware image formation model for more accurate defocus blur at depth discontinuities[91], hyperspectral depth estimation[92], polarization-multiplexed modulation for better depth cues[93] as well as color-coded apertures[94].

Another computational imaging paradigm tackled by 'deeply coded apertures' is full-spectrum and hyperspectral imaging[95–100]. Joint optimization of a coded aperture and a reconstruction network has been incorporated into the coded aperture snapshot spectral imager (CASSI) framework[95,100], which uses a binary-coded aperture and dispersive elements such as a prism in addition to the imaging lens for introducing spectral variance. Recent works[96–98] proposed a single diffractive optical element to encode the spectral information, resulting in compact hardware. Jeon et al.[96] crafted a DOE design that generates a set of spectrally varying anisotropic PSFs, and subsequently trained neural networks to reconstruct the spectral images from the sensor readout through the DOE. In contrast, as depicted in Fig. 2a, Dun et al.[97] employed an end-to-end design framework to learn a diffractive achromat (DA) for high-fidelity full-spectrum imaging together with the reconstruction CNN. They used concentric ring decomposition to learn a rotationally symmetric diffractive achromat design, reducing computational complexity and memory demand by an order of magnitude. Compared to the DOE designed by following the sequential approach in Ref. [96], they reported a 1.3 dB improvement in the peak signal-to-noise ratio of the recovered images. This demonstrates that optimizing an aperture jointly with the reconstruction ANN in a supervised manner outperforms carefully handcrafted aperture designs working in conjunction with a separately trained ANN.

Deeply coded apertures have also been utilized for lens-free imaging[86,101], super-resolution imaging[102], extended depth-of-field[103,104], and high dynamic range (HDR) imaging[105,106]. Figures 1 and 2 illustrate some of these examples. Other applications of deeply coded apertures include light-field photography[107–109], video compressive sensing[110–112], and localization microscopy[113–115], some of which are depicted in Fig. 3. For example, Higham et al.[110] trained a convolutional auto-encoder network for 'real-time single-pixel video', i.e., video-rate image reconstruction with a single-pixel camera. In their implementation, the measurement bases for the image intensities are learned as the first layer of an auto-encoder network. Measurement with these 'deeply learned' bases is physically executed using a high-speed digital micromirror device (DMD), as shown in Fig. 3a. Operating at a DMD modulation rate of 20kHz, they were able to recover 128×128-pixel images from 333 single-pixel measurements at 30 frames-per-second, obtaining a compression of ~50-fold. Inagaki et al.[107] used end-to-end optimization of coded apertures and ANNs for capturing light fields by modeling the front-end optics as the encoder and the back-end ANN as the decoder within an auto-encoder architecture. They demonstrated reconstruction of 5×5 or 8×8 viewpoints of 64×64-pixel images from 1, 2, or 4 measurements while showing the advantage of ANN-assisted reconstruction compared to previous methods (Fig. 3b).

As another example, Hershko et al.[113] demonstrated multicolor localization microscopy with PSF engineering using an SLM at the Fourier plane of the imaging system, as depicted in Fig. 3c. The SLM voltage pattern, resulting in distinct PSFs at distinct wavelengths for maximal color distinguishability, was optimized in conjunction with a reconstruction ANN. The reconstruction network ultimately outputs the position and the color of the emitters in the image. In a later work, the researchers also demonstrated 3D localization microscopy enabled by PSF engineering, following a similar principle[114]. Another microscopic modality where engineered diffraction and digital reconstruction networks have been shown to work hand-in-hand is ptychography[116–120], where the illumination patterns aimed for optimal coding are learned by embedding them within the reconstruction network. This has been carried out for diverse tasks such as image classification, quantitative phase imaging (QPI), super-resolution imaging, etc. Figure 3d depicts QPI enabled by physics-based learning of coded illumination patterns[119].

Diffractive optical networks have also been used for computational imaging and sensing applications[69,71,74,75,79,121–126]. Several of these earlier works fall outside the scope of this article as they utilized supervised learning and deep learning-based optimization for engineering a diffractive processor to perform all-optical image reconstruction, such as all-optical QPI, complex field imaging, seeing through random diffusers and class-specific imaging, among others. There are also examples of diffractive optical networks that are jointly trained with ANNs to perform computational imaging and image reconstruction in a hybrid (optical + digital) format, some of which will be covered in the following sections[65,68,76,127–129].

## Reconfigurable Applications of Programmable Diffraction and ANNs

The collaboration of engineered diffraction with digital ANNs has been used in the field of meta-optics[56], i.e., the branch of optics dealing with the design and applications of metamaterials and metasurfaces. The utility of metasurfaces for many applications demands reconfigurability to address real-world scenarios. A reconfigurable metasurface, also known as a programmable metasurface, is a type of metasurface that offers dynamic and programmable control over the manipulation of electromagnetic waves. Unlike traditional metasurfaces, which have fixed geometries and properties, reconfigurable metasurfaces employ an array of individually addressable elements that can be controlled to dynamically change the response of the surface to incident waves. However, finding the control code to program a reconfigurable metasurface for a desired response poses an inverse problem, the solution of which is hard to obtain in real-time using conventional optimization algorithms. The use of deep learning has enabled real-time solutions to such inverse problems, where ANNs were utilized for rapid prediction of the metasurface code for a desired response, bringing the potential of reconfigurable metasurfaces to reality for various real-world applications[130,131]. For example, Li et al.[132] reported ANN-enhanced application of dynamic metasurfaces and demonstrated a smart metasurface imager operating at 2.4 GHz, powered by adaptive control of dataflow using ANNs, see Fig. 4a. They used a CNN (IM-CNN-1) to reconstruct a high-resolution image of the human body from microwave data; in this mode, the metasurface serves as a coded aperture for compressive imaging. Another CNN (R-CNN) was used to find the region of interest (ROI) within the image, e.g., the chest for respiration monitoring or the hand for hand-gesture recognition. Finally, another CNN (IM-CNN-2) processed the microwave data reflected from the ROI to, for example, recognize hand signs.

As another example, the use of deep learning for predicting metasurface codes was demonstrated by Shan et al.[133]. Subsequently, Li et al.[134] took a physics-inspired approach to train an ANN to dynamically predict the required metasurface code for intelligent beamforming. Similarly, Li et al.[135] incorporated deep learning into compressive imaging using a programmable metasurface aperture. However, instead of utilizing ANNs only for image reconstruction, they jointly optimized the control coding patterns of the metasurface also as part of an ANN. Similar strategies were followed in other works[136,137]. Li et al.[138] also demonstrated smart beam tracking and wireless communication with a moving target utilizing reconfigurable metasurfaces and ANNs. They employed a CNN, together with a depth camera (mounted on a programmable metasurface, see Fig. 4b), to automatically detect the position of the moving target in the environment. Another trained ANN took this position information as input and predicted the coding sequence (in a few milliseconds) to drive the metasurface for intelligent beam tracking through a field-programmable gate array (FPGA). Hu et al.[139] also provided a theoretical analysis of deep reinforcement learning for predicting the codes of a beamforming metasurface used for the 3D localization of objects (Fig. 4c).

As another application, Qian et al.[140] demonstrated intelligent cloaking with deep learning-enabled metasurfaces. They realized intelligent cloaking with a rapid response to fluctuating incident waves and surrounding conditions, all without human intervention, utilizing a few key components: a reconfigurable metasurface, two detectors, and a trained ANN (see Fig. 4d). The active meta-atoms within the ultrathin metasurface create spatially varying local reflections to generate a back-scattered wave that is characteristic of the surroundings in the absence of the cloaked object. The ANN was trained to predict the meta-atoms' DC bias voltages required for the desired local reflection phases, based on input from the surrounding environment and the incident waves. Experimental demonstrations confirmed the functionality of this self-adaptive invisibility cloak in the microwave domain, with potential scalability to higher frequencies. In other works, deep learning was also leveraged to realize cloaking and optical illusion with conformal metasurfaces.[141–143]

## Hybrid (Optical-Electronic) Information Encoding and Decoding

Diffractive processing can also be exploited to accelerate ANN computations. Fast training and execution of deep ANNs are made possible through the adoption of dedicated digital hardware platforms such as Graphics Processing Units (GPUs), Tensor Processing Units (TPUs), etc.[144] However, the execution of ANNs, with the number of parameters easily exceeding tens to hundreds of millions, is power-intensive in digital electronic platforms. The bandwidth of digital electronics is also insufficient to allow for substantial acceleration of inference speed with such ANNs. At the same time, there has been an apparent stagnation in Moore's law over the past decade, which has driven the miniaturization of transistors, the key component for digital electronics, since the seventies. These limitations have inspired researchers to look for a suitable alternative to digital electronics for the implementation of ANNs[145]. One such alternative is optics/photonics with its large bandwidth and massive parallelism available for fast and high-throughput computing.[146] Optical implementations of neural networks can potentially achieve faster processing speed with lower energy consumption, increased bandwidth, and improved scalability. One limitation of the optical computing paradigm, however, is the lack of ultra-fast, low-power, low-loss, and broadband nonlinear operations in a cost-effective and compact format. Hence, the possibility of realizing a general-purpose computer or even a neural network in an exclusively optical platform faces challenges. Therefore, one can argue that the sweet spot for operation could be at the intersection of optics and electronics. One solution for accelerating neural network performance could be executing the front-end linear operations in the optical domain, offloading these computationally expensive operations to low-power and high-speed optics. For example, to meet the increasing need for computing resources, integrated photonic circuit-based accelerators have been proposed to improve ANN inference, particularly focusing on accelerating the execution of linear operations such as matrix multiplications[147]. However, such efforts are excluded from this article's coverage as we only consider free-space optics-based neural network accelerators that exploit engineered diffraction. As an example of the latter, one area where such a collaboration between engineered free-space diffraction and ANNs has been used for statistical inference and machine learning tasks is *compressive classification*, where optically coded low-resolution sensor readouts are used for image classification by a jointly-trained back-end neural network, enabling machine vision even with a single-pixel sensor[121,127]. Apart from reducing the number of digital operations following optoelectronic detection, such a compressive classification scheme enabled by the collaboration between engineered diffraction and ANNs also lowers the requirement on the number of pixels at the optics-electronics interface, increasing the frame rate while reducing power and bandwidth requirements. For example, Mengu et. al.[65] explored the integration of diffractive networks with digital neural networks for statistical inference, where the compressed detector readout of the diffractive network output was fed to a shallow ANN to limit the number of digital operations (Fig. 5d). In this compressive image classification paradigm, the diffractive network all-optically encodes the object information to a lower-dimensional representation, which is subsequently processed by a shallow digital ANN for lower-power classification of input objects. This framework obviates the need for a 4f system comprising bulky lenses and forms a super-set to such systems that can only perform convolutional filters, whereas a thin diffractive processor can optically perform any fully connected complex-valued linear operation at the diffraction limit of light, within an axial thickness of a few hundred wavelengths. Diffractive networks were also used for encoding the input object information into the spectrum of a broadband illumination, which was read out using a time-resolved single-pixel detector and subsequently used for inferring the object class as well as reconstructing its image from a highly compressed spectral representation[127]. This single-pixel image classification and reconstruction performance was enabled by space-to-spectral encoding performed through a diffractive optical network (Fig. 5e). Similarly, Bacca et al.[148] simultaneously optimized the binary sensing matrices of a single-pixel camera and the parameters of a classification ANN such that the obtained single-pixel measurements can be used for object classification. For another application, del Hougne et al.[149] used the joint optimization of the acquisition hardware, i.e., reconfigurable metasurface apertures of a transceiver (see Fig. 5a), and postprocessing software, i.e., a neural network, for object recognition with a limited number of measurements from a single-pixel detector. As another example of the collaboration between engineered diffraction and ANNs, Chang et al.[150] optimized the phase masks at the Fourier plane of a 4f system to optically implement the convolution kernels of the first layer of a neural network, as shown in Fig. 5b. They used tiling on the sensor plane to implement different output channels and the sensor readout was then fed into the rest of the network with the necessary preprocessing of data. Similarly, Zheng et al.[151] used metasurface-based imagers designed to work together with a digital back-end neural network for angle and polarization-multiplexed multichannel front-end optical information processing (Fig. 5c).

Another exciting opportunity enabled by the collaboration of engineered diffraction and digital neural networks is the complementary encoding and decoding of information. Such an encoder-decoder-based collaboration might be aimed at bypassing a physical bottleneck in the system. For example, Isil et al.[68] demonstrated pixel super-resolution (SR) holographic display, where a low-pixel count SLM can be used to display a high-resolution image. In this framework, a digital CNN predicts the low-resolution SLM phase pattern corresponding to a high-resolution image. The transmitted phase pattern is all-optically decoded by a jointly-trained diffractive network to display the high-resolution image at the output plane, see Fig. 6a. Rahman et al.[128] exploited a similar collaboration of a digital ANN and a jointly-optimized diffractive network to transfer any arbitrary image/message around fully opaque, zero-transmittance occlusions that are dynamically changing (Fig. 6b). Here, for an arbitrary image/message to be transferred, the encoded phase pattern that is to be transmitted is predicted by a digital CNN, and the transmitted wave, after getting blocked and scattered by the opaque occlusion, arrives at the receiver, where a diffractive network all-optically decodes the wave and projects the target image/message at the output, see Fig. 6b. This scheme has been shown to work even when all the direct path of light rays between the transmitter and the receiver are blocked by the occlusion, albeit with poor diffraction efficiency due to the weak scattering from the occlusion edges available to carry information. A similar strategy has also been used for transferring arbitrary images/messages through random unknown diffusing media,[129] see Fig. 6c.

There are also applications where the roles of the encoder and the decoder are swapped between the electronic network and the diffractive network, such as an information-hiding camera[76], as shown in Fig. 6d. Here, the diffractive encoder all-optically conceals any input image/message of interest into ordinary-looking patterns (deceiving human observers to believe that the object/message is something else), which can be transmitted without raising any alarm on the part of a potential eavesdropper. These ordinary-looking images captured by the information-hiding camera, however, can be accurately and rapidly deciphered by an electronic decoder to reveal the original information.

## Conclusions and Prospects

We surveyed the myriad of exciting applications that leverage the collaboration between engineered diffraction and digital ANNs. For deep neural networks, sizeable savings in terms of computational resources have been attained by implementing the front-end of the network with engineered diffractive optics. For even larger gains, this capability needs to be extended to the implementation of more challenging tasks, such as classification over a large number of data classes. This opportunity also hinges upon the realization of suitable optical nonlinearities. There have been demonstrations where each layer of diffractive linear processing is interleaved by a digital nonlinear activation function[63]; however, the series of conversions between the optical and electronic domains required by such implementations partially overshadow the advantages of optical computing, such as its speed and power-efficiency. To reap the maximum benefits of engineered diffraction, optical nonlinearities that are low-power, ultra-fast, broadband and low-loss would be needed[152], ideally in a compact and cost-effective format that is scalable for large-volume manufacturing.

For computational imaging and sensing applications, deeply coded apertures have been shown to provide better performance than alternative methods. However, these applications, in general, still use larger ANN back-ends, which might hinder their adoption in edge devices and resource-limited settings. This aspect could potentially be addressed by innovations both in the optical and digital domains, such as better utilization of the information capacity of optics by exploiting the relevant degrees of freedom as well as miniaturization of the size of the ANNs, making them shallower and lower power.

In nearly all cases explored in the literature, metasurface-based applications use plasmonic metasurfaces for their reconfigurability, which are relatively lossy. Also, these plasmonic metasurfaces are addressed by digital electronics with limited bandwidth, which could be a bottleneck in their operating speed. Therefore, the search for alternatives, such as dielectric reconfigurable metasurfaces[153–157], is important for energy-efficient and faster implementations.

Furthermore, the rapid progress in 3D fabrication technologies continues to enable the construction of diffractive features with sub-wavelength dimensions, appropriate for visible and IR applications. This would potentially result in increased computing density (at the diffraction limit of light) attainable with diffraction engineering via the miniaturization of diffractive devices. Consequently, integration of free-space diffractive computing with ANNs could serve low-power, multi-wavelength, high-speed and high-throughput computing while maintaining small footprints.

Finally, reconfigurable diffractive optics powered by digital ANNs have already unlocked a variety of intriguing applications. There are many other potential applications that can be revolutionized by the integration of digital ANNs with diffractive computing, such as edge computing on mobile platforms and augmented- and virtual-reality display technologies, a front that has gained extreme interest in the past few years. Overall, future research advances in combining engineered diffraction of light with digital ANNs hold the potential to transform many disciplines by unlocking new capabilities in computation, communication, and visual information processing.

**Figures and captions**:

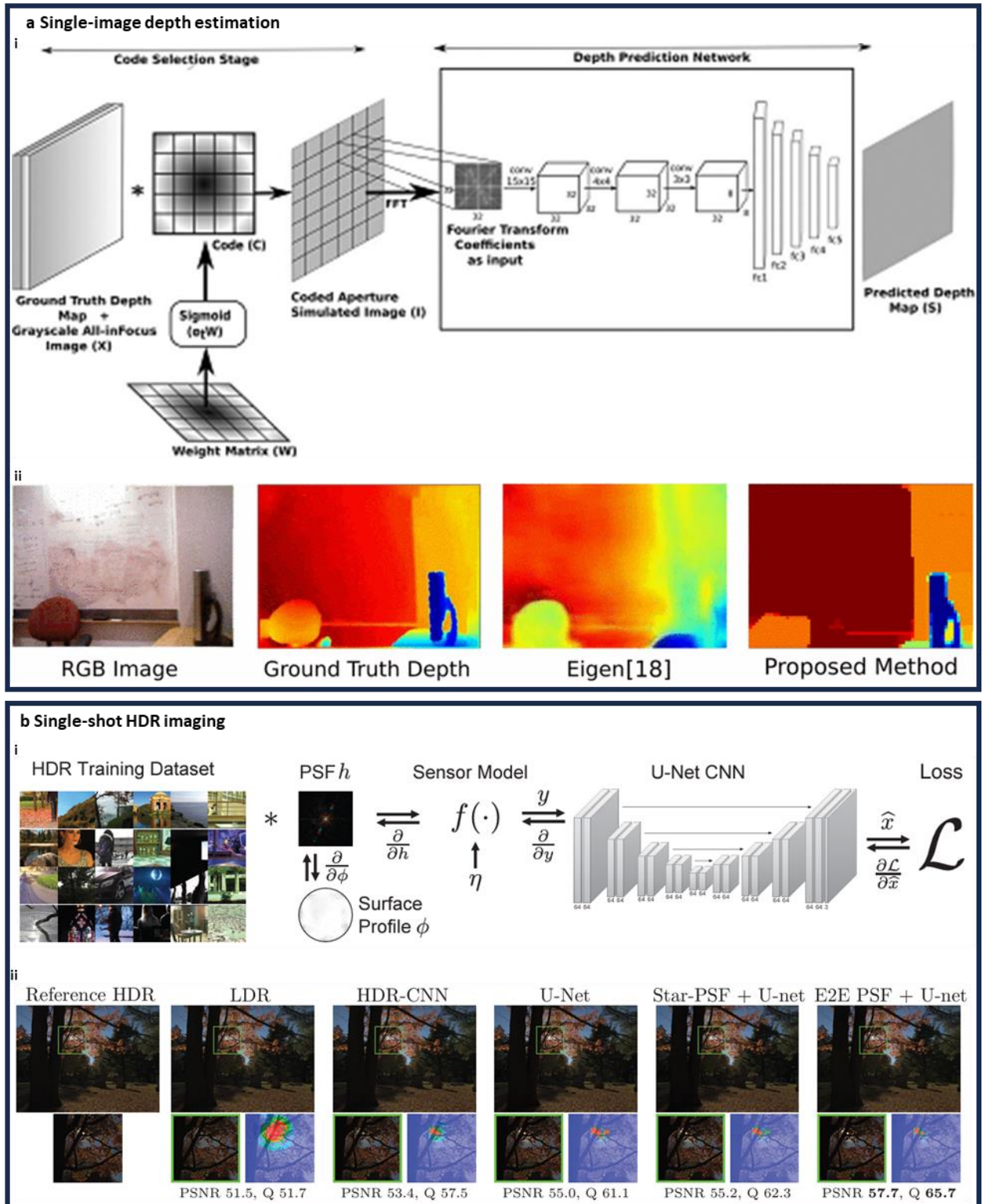

**Figure 1**. Deeply coded aperture for computational imaging. (a) Single-image depth estimation. i. For end-to-end training, the effect of the coded aperture is embedded within the deep learning framework

by convolving the ground truth image with the aperture code; to enforce binarization, the code is obtained by applying the sigmoid function to a learnable weight matrix. The coded image goes through the depth prediction network. Error in depth map prediction is backpropagated to the network weights and the code weight matrix. ii. Comparison of the proposed method with a prior method based solely on a CNN without any deeply coded aperture. Adapted from Ref [88]. (b) Single-shot HDR imaging. i. HDR images from a training set are convolved with the PSF corresponding to a lens surface profile denoted by $\phi$. The function $f(\cdot)$ models the sensor saturation, and $\eta$ denotes the additive sensor noise. A CNN processes the resulting sensor measurement and outputs an estimate of the HDR image. The loss between the ground truth and the estimated HDR image is propagated backward to optimize both the CNN parameters and the height values ($\phi$) of the lens surface in an end-to-end manner. ii. The end-to-end design approach outperforms other approaches both qualitatively and quantitatively. Adapted from Ref [105].

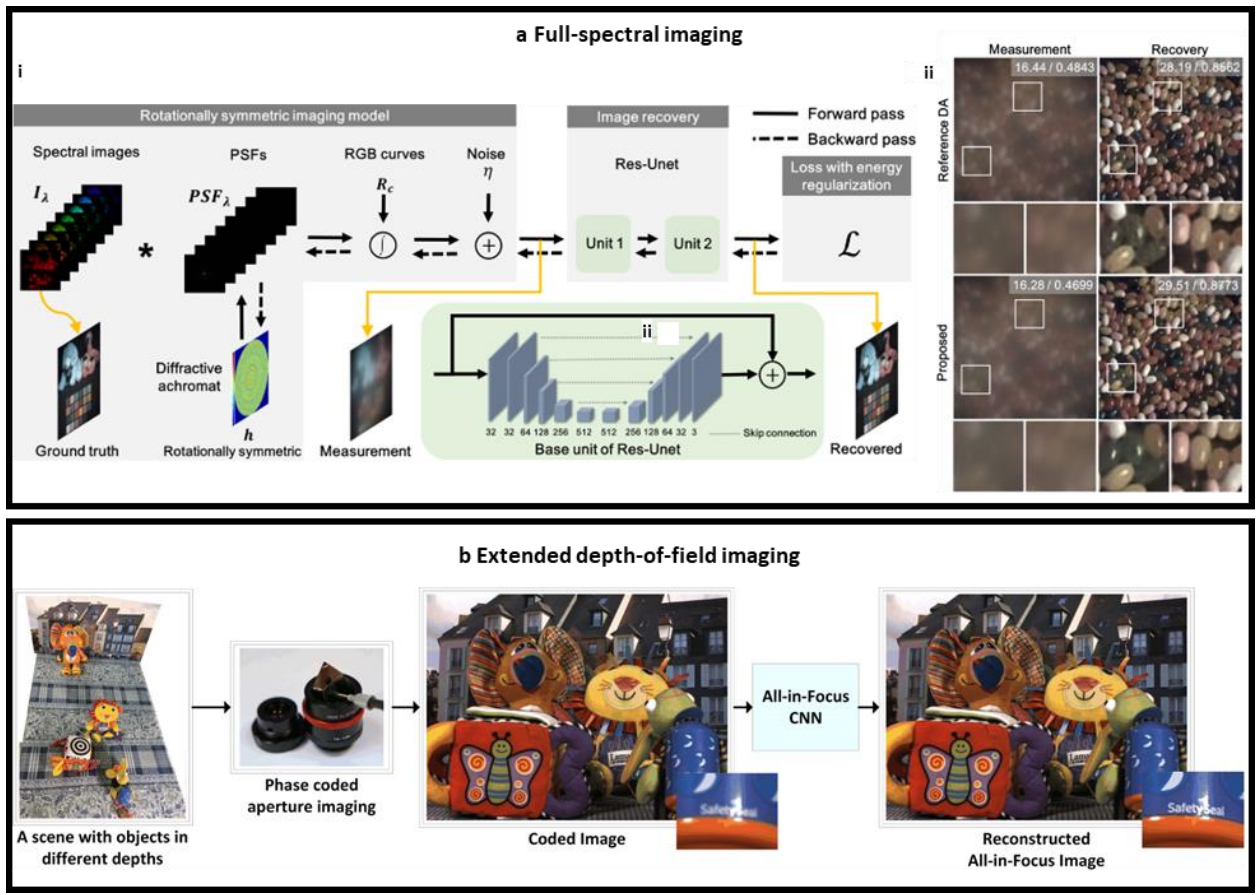

**Figure 2**. Deeply coded aperture for computational imaging. (a) Full-spectral imaging with an RGB sensor. i. Design of a diffractive achromat for full-spectral imaging. The procedure is similar to the one shown in Fig. 1b, except that the ground truth spectral images are convolved by the PSFs of the DA at the respective wavelengths, and the RGB sensor measurement is simulated by integrating over the color response of the sensor for each channel. ii. Performance comparison reveals better recovery for the DA designed using the proposed approach compared against a reference DA. Adapted from Ref [97]. (b)

Extended depth-of-field imaging. A coded aperture camera captures the coded image, whereupon a CNN reconstructs the all-in-focus image. Adapted from Ref [103].

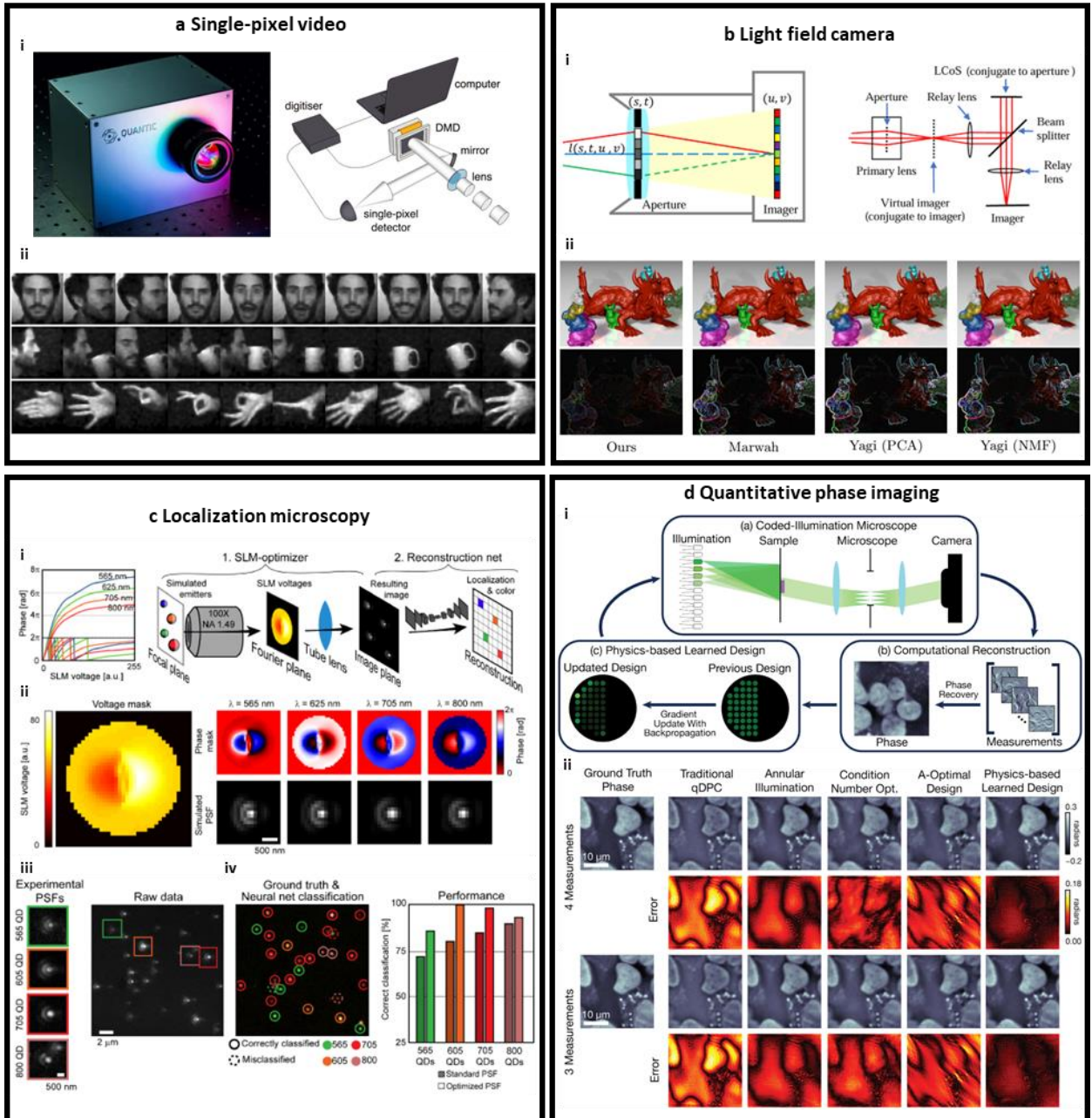

**Figure 3.** Deeply coded aperture or illumination for computational imaging. (a) Single-pixel video. i. A single-pixel camera system for video-rate image reconstruction, with a simplified schematic on the right. The lens forms an image of the scene onto the DMD, which modulates the light using a series of $M$ binary patterns (128 × 128 pixels) that are optimized together with a CNN to reconstruct a frame from the $M$ single-pixel measurements. ii. Selected frames (128 × 128 pixels) from single-pixel camera videos with $M = 666$ (4% of the number of pixels). For $M = 666$, the frame rate is 15 Hz. Adapted from Ref [110]. (b) Light field photography. i. Schematic and implementation of coded aperture camera for light field

photography. ii. Reconstructed image obtained with a deeply coded aperture design and the difference from the ground truth, compared with other compressive sensing-based methods. Among the $M = 5 \times 5$ reconstructed views using $N = 2$ acquired images, only the central view is shown for comparison. Adapted from Ref [107]. (c) Multicolor localization microscopy. i. PSF engineering through an SLM placed at the exit aperture of a microscope objective for emitter localization. The SLM voltages are optimized in conjunction with a back-end digital reconstruction network in an end-to-end manner. ii. The optimized SLM voltage pattern, the phase delay imparted to 565, 625, 705 and 800 nm light at the optimized SLM voltages and the corresponding simulated PSFs. iii. Experimental PSFs obtained for four quantum dots (QDs) within a larger FOV. iv. Color classification of QDs within the same FOV, denoted by circles overlaying the emitters that are artificially colored according to their ground truth wavelengths. A comparison of performance in terms of color classification accuracies with the standard and optimized PSFs is also shown. Adapted from Ref [113]. (d) Quantitative phase imaging (QPI). i. The LED-array microscope captures multiple intensity measurements with different coded illumination patterns, which are used to computationally reconstruct the sample's complex field. The computational reconstruction utilizes a physics-based unrolled network, which reduces the required number of training examples. ii. Performance comparison with other methods reveals lower phase reconstruction error with the physics-based learned design. Adapted from Ref [119].

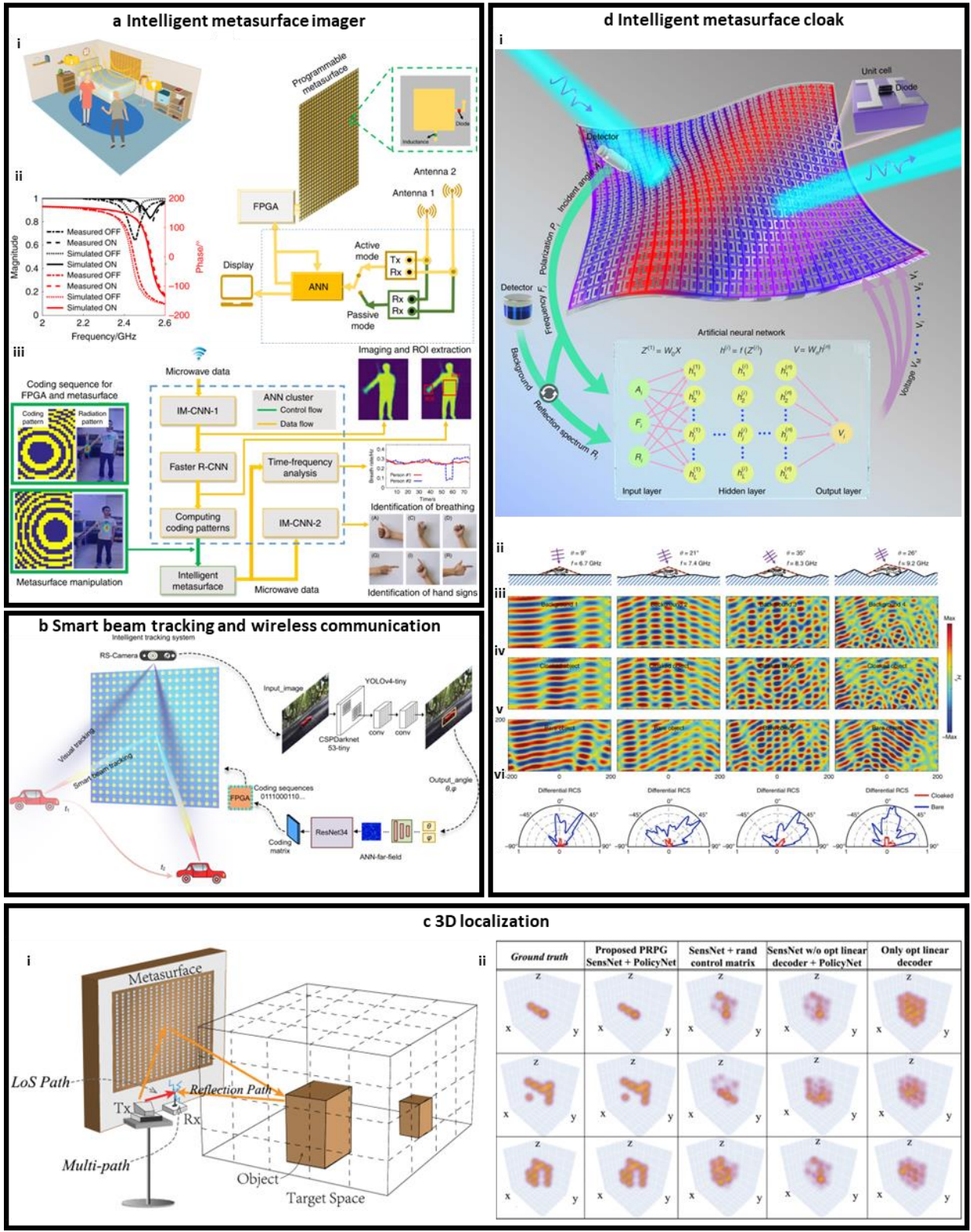

**Figure 4.** Intelligent reconfigurable metasurfaces. (a) Intelligent metasurface imager. i. The conceptual scenario where a metasurface decorated as part of a wall intelligently scans the environment for relevant

information. In the passive mode, the large aperture metasurface, together with the antennas, is used for high-resolution imaging in a compressive sensing manner. In the active mode, the metasurface is used to focus microwave radiation from the transmitter antenna to the region of interest within the environment. Active beamforming is attained by reconfiguring the metasurface through the voltages applied to the PIN diodes loaded onto the meta-atoms. ii. Experimental and simulated frequency responses of the designed meta-atom at different PIN diode states. iii. Data flow within the framework. Microwave data from the antennas are processed first with a CNN (IM-CNN-1) to form the image of the whole scene and then with another faster CNN (R-CNN) to find the ROI. The ROI can either be the chest for respiration monitoring or the hand for sign-language recognition. Subsequently, the metasurface is configured to focus the transmitted radiation toward the desired ROI. The back-reflected radiation is either processed by another CNN (IM-CNN-2) to recognize the hand sign or time-frequency analyzed for respiratory monitoring. Adapted from Ref [132]. (b) Smart beam tracking for wireless communication. The depth camera (RS-camera) independently detects the position of a moving target with the help of a target-detection CNN. This positional information of the specified target is used as input to a pre-trained ANN (ANN-far-field). In a matter of milliseconds, the ANN produces a coding sequence for the metasurface, which is subsequently sent to the programmable metasurface through an FPGA, enabling smart beam tracking for enhanced wireless communication capabilities. Adapted from Ref [138]. (c) 3D localization. i. Representation of the compressive RF sensing scenario aided by metasurfaces: The signals emitted by the transmitter (Tx) unit undergo modulation upon reflection from the metasurface before entering the target space. These signals are then reflected by objects within the target space and captured by the receiver (Rx) unit. A deep reinforcement learning algorithm interprets these received signals to detect the presence or absence of objects at each spatial grid within the target space. ii. A comparison between the ground truth and the sensing outcomes of objects with various shapes, evaluated against different benchmark algorithms, is depicted. Adapted from Ref [139]. (d) Dynamic cloaking with a metasurface. i. A reconfigurable metasurface equipped with two detectors and a power supply dynamically cloaks an underlying object in response to a dynamically changing environment. Upon detection of an incoming wave and information regarding the surrounding background, a trained ANN automatically computes the required bias voltage of each meta-atom for active cloaking. ii. Near-field magnetic field distributions of the background (iii), cloaked object (iv) and bare object (v), as well as far-field differential radar cross-section (RCS) (vi), for the four settings of the background and the incident wave shown in (ii). Adapted from Ref [140].

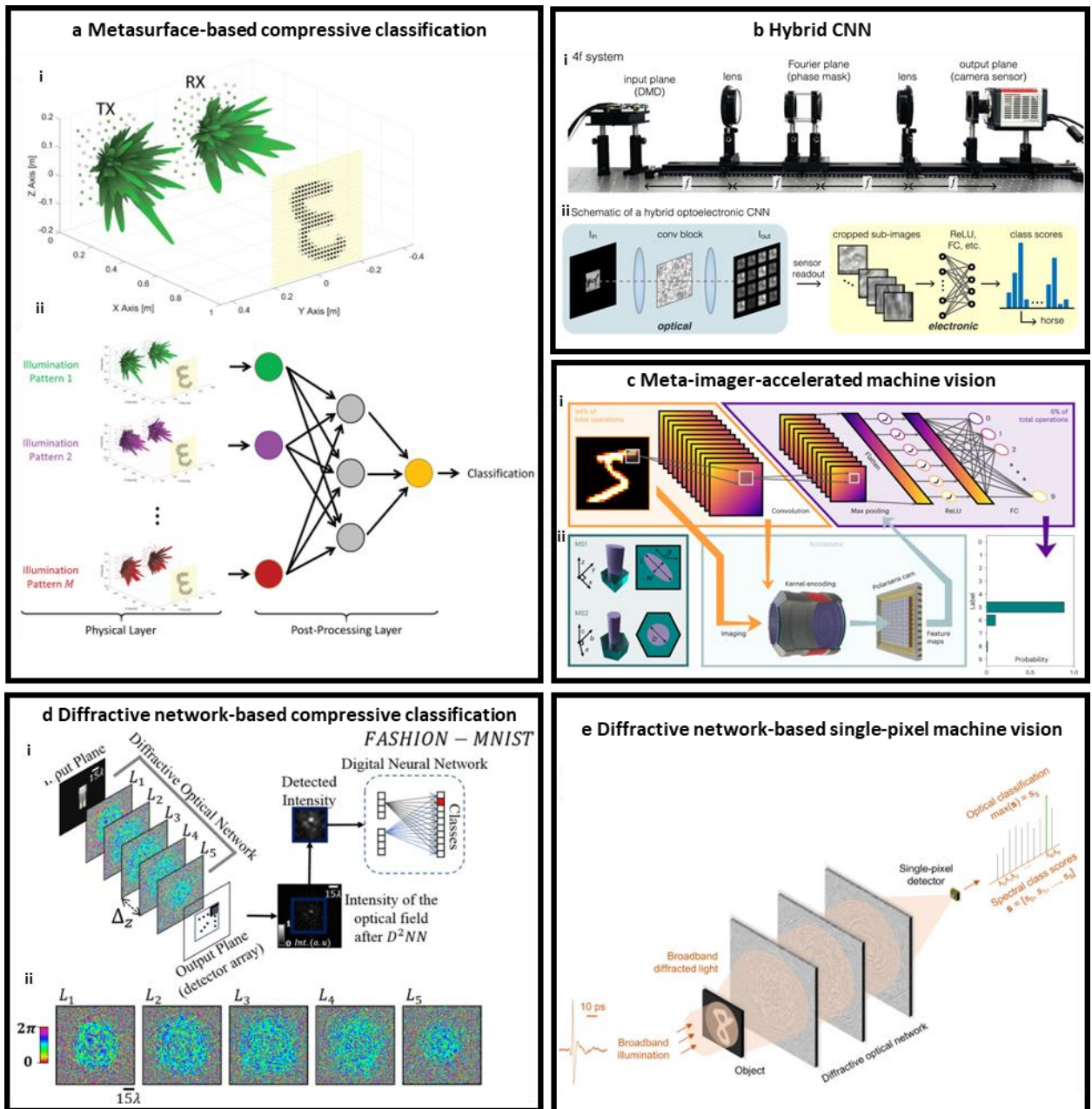

**Figure 5.** Engineered diffractive computing for statistical inference. (a) Compressive classification based on a metasurface aperture. i. The framework entails actively illuminating the scene with a transmitter (TX) metasurface and capturing the reflected waves with a separate receiver (RX) metasurface. ii. The scene is illuminated using M different TX-RX metasurface configurations, resulting in a complex-valued measurement vector of size M. The measurement is then processed by an artificial neural network, which outputs a classification score of the scene. Adapted from Ref [149]. (b) Hybrid optical-electronic CNN. i. Implementation of a convolutional layer with optics. A 4f system can be adapted to all-optically compute the feature maps of a convolutional layer by integrating a phase mask into the Fourier plane. ii. A hybrid CNN where the first convolutional layer is implemented optically. The feature maps corresponding to the optimized convolutional kernels are arranged as sub-images on a 2D grid on the sensor. The sensor readout is processed, and the cropped sub-images are fed into subsequent digital

layers. Adapted from Ref [150]. (c) Meta-imager-accelerated machine vision. i. The convolution operations of a shallow CNN with twelve output channels are carried out using an angle- and polarization-multiplexed meta-imager. The resulting feature maps are captured by a polarization-sensitive camera and subsequently fed into a trained digital neural network to produce a probability histogram for image classification. ii. The convolutional kernels are encoded into two cascaded metasurfaces within the meta-imager. The schematic of the meta-atoms for the first (MS1) and second (MS2) metasurfaces are shown. Adapted from Ref [151]. (d) Hybrid classification utilizing a diffractive optical network. i. The architecture of a hybrid classifier, which combines both optical and electronic processing. A diffractive network performs optical image encoding for the purpose of subsequent digital classification on the compressed object information by a shallow ANN. A 10×10 optoelectronic detector array positioned after the diffractive network front-end serves as the interface with a shallow fully connected ANN back-end. ii. The optimized phase-only optical layers that are positioned at the forefront of a hybrid fashion product classifier. Adapted from Ref [65]. (e) Spectrally encoded single-pixel machine vision. A diffractive optical network encodes the spatial information of input objects, which are illuminated by a broadband source, into the spectrum of the output light captured by a single-pixel detector. The class of each input object is determined by the highest spectral intensity, where the spectral intensities are measured at a set of discrete wavelengths, each corresponding to a distinct data class. Adapted from Ref [127].

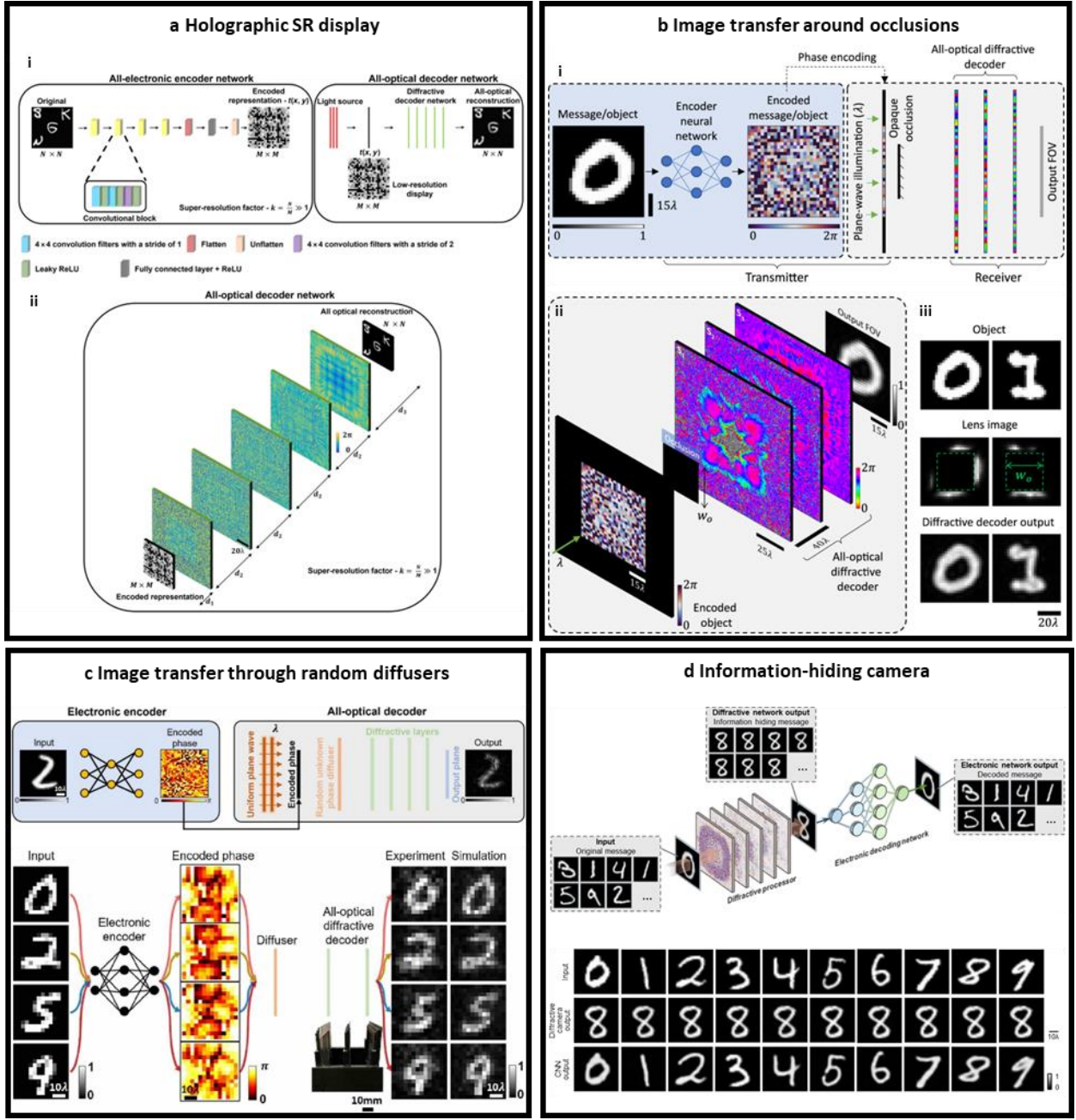

**Figure 6**. Hybrid optical-electronic information encoding and decoding. (a) Holographic SR display. i. An electronic encoder ANN generates low-resolution codes of high-resolution input images, which are projected by a low pixel-count phase SLM. These low-resolution projected fields are subsequently decoded all-optically by the diffractive network, displaying the high-resolution images at the output FOV. ii. Illustrative 3D layout of the five-layer diffractive network decoder. Adapted from Ref [68]. (b) Image transfer around an arbitrary opaque occlusion that is dynamically changing. i. At the transmitter, a phase-coded representation of the image/message to be transferred is generated by a digital ANN. The encoding is aimed at bypassing the opaque occlusion. The transmitted phase-coded wave gets blocked and scattered by the occlusion before arriving at the receiver, where it is all-optically decoded by a diffractive network that is jointly trained to recover the input image/message. ii. 3D illustration of the transmitter aperture, occlusion, and diffractive network decoder. iii. The electronic encoding-diffractive decoding scheme recovers the image information lost by a lens-based imaging system in the presence of an opaque occlusion. Adapted from Ref [128]. (c) Image transfer through random unknown diffusers. An electronic encoder generates a phase-encoded representation of the image to be transferred through random unknown phase diffusers. The transmitted phase-encoded wave, after being perturbed by the diffuser, is all-optically decoded by a diffractive network. Adapted from Ref [158] (d) Information hiding camera. A diffractive network all-optically conceals the input optical information/message into ordinary-looking output images that deceive human observers but can be deciphered by an electronic decoder network to reveal the original information/message, boosting visual information security. Adapted from Ref [76].